\documentstyle[12pt]{article}
 1

\begin{document}
\begin{flushright}
{IOA.303/94}\\
NTUA 44/94\\
\end{flushright}
\begin{center}
{\bf  IMPLICATIONS OF A HEAVY TOP}\\
{\bf    IN SUPERSYMMETRIC THEORIES}\\
\vspace*{1cm}
{\bf G.K. Leontaris}\\
{\it Theoretical Physics Division} \\
{\it Ioannina University} \\
{\it GR-45110 Greece} \\
\vspace*{0.3cm}
{\bf and}\\
\vspace*{0.3cm}
{\bf N.D. Tracas}\\
{\it Physics Department}\\
{\it National Technical University}\\
{\it GR-157 80 Zografou, Athens, Greece}\\
\vspace*{0.5cm}
{\bf ABSTRACT} \\
\end{center}
\noindent
In the context of the radiative electroweak symmetry breaking
scenario,  we investigate the implications of a heavy top quark mass,
close to its infrared fixed point, on the low energy parameters of the
minimal supersymmetric standard model. We use analytic expressions to
calculate the Higgs  masses as well as the supersymmetric masses  of
the third generation. We further assume bottom-tau unification at the
GUT scale and examine the constraints put by this condition on the
parameter space ($\tan\beta$,$\alpha_3$), using the renormalization
group procedure at the two-loop level. We find only a small fraction of
the parameter space where the above conditions can be satisfied, namely
$1\le \tan\beta \le 2$, while $0.111\le\alpha_3(M_Z) \le 0.118$.
We further analyse the case where all three
Yukawa couplings reach the perturbative limit just after the
unification scale. In this latter case, the situation turns out to be very
strict demanding $\tan\beta\sim 63$.

\vspace*{2cm}
\noindent
\begin{flushleft}
IOA-303/94\\
NTUA 44/94\\
March 1994
\end{flushleft}
\thispagestyle{empty}
\vfill\eject
\setcounter{page}{1}


\noindent
{\bf Introduction}

The last years there has been a revived interest in the supergravity
unified models and their low energy effective theories, mainly due to
the fact that LEP measurements are in good agreement with a gauge
coupling constant unification scenario with supersymmetric
$\beta$-function coefficients down to the scale of  $\sim 1$TeV.
However, the existence of supersymmetry will only be confirmed when
new particles  --  the superpartners of the standard model spectrum --
will be observed in (near) future experiments. Thus, the study of
supersymmetric grand unification is very important and should be seen
in conjuction with the predictions for the new particles which may be
observed soon. So far, the constraints put by the unification of the
three gauge couplings require a superpartner mass spectrum in the
range of $(0.1-1)$TeV which can be accessible in the near future.

There is another experimental fact the last few years which seems to
be related with the fate of the electroweak symmetry breaking  in an
effective supersymmetric low energy theory. The non-observation of the
top quark gives a lower bound on its mass $m_t
\stackrel{>}{\sim}100$GeV. Although this result is disappointing from
the experimental point of view, on the other hand, it fits perfectly
with the idea of radiative symmetry breaking scenario suggested
several years ago
$^{\cite{IR,RAD}}$.
Indeed the
renormalization group impoved SUSY Higgs potential breaks the
$[SU(2)\times U(1)]_{EW}$ symmetry when the top Yukawa coupling is
large enough  to drive one of the soft supersymmetry breaking
parameters (namely $m_{H_2}^2$) negative.

Grand unification based on the most popular groups, with the minimal
number of fermion and Higgs content, implies additional relations in
the initial values of the parameters of the theory. Thus, for example
in the Yukawa sector, one such well known constraint requires the
bottom and tau lepton Yukawa couplings $h_b$ and $h_{\tau}$, to be
equal at the unification scale $E_G$
\begin{equation} h_{b}(E_G)=h_{\tau}(E_G)  \label{eq:1} \end{equation}
In certain cases, and particularly in string derived unified models,
additional constraints on the Yukawa sector of the theory are often
obtained, i.e.
\begin{equation} h_{b}(E_G)\approx h_{\tau}(E_G)\approx h_{t}(E_G)\sim
g_{String}\label{eq:2} \end{equation}
where $h_t$ is the top Yukawa coupling and
$g_{String}$  is the value of the unified gauge coupling at the
string scale $E_{String}\ge E_G\sim 10^{16}$GeV.
In patricular, a large top Yukawa coupling which is implied by the last
equality in Eq.(\ref{eq:2}), motivates again the study of the
fixed point solutions proposed several years ago  in the context of non
supersymmetric theories
$^{\cite{PR}}$.

Motivated by the experimental fact that the top quark mass is rather high
as well as from the aforementioned theoretical speculations, in the present
work we wish to study the implications of the above considerations on the low
energy theory.
In order to minimize the arbitrary parameters and to avoid complications with
flavour changing neutral currents, we assume universality of the scalar mass
parameters at the GUT scale.
Using renormalization group techniques, we derive the mass formulae of the
scalar masses (in particular those affected by a large top quark mass)  and
examine their properties close to the infrared fixed point of the top mass.
Furthermore we investigate the regions of the parameter
$\tan\beta =<H_2>/<H_1>$ which
are compatible with the above constraints and the minimization conditions put
by the renormalization group improved Higgs potential.


\noindent
{\bf Radiative Symmetry Breaking in the Presence of a Heavy Top Quark }

One of the most appealing features of supergravity theories is the
radiative symmetry breaking mechanism
$^{\cite{RAD}}$
which may occur in the presence
of a heavy top quark mass.  Indeed, the renormalization group improved
Higgs potential breaks the electroweak symmetry if the top Yukawa
coupling is large enough to drive the $m_{H_2}^2$ mass parameter
negative below a certain scale $Q_0$.

At the tree level the supersymmetric Higgs potential can be written as
follows
\begin{eqnarray}
{\cal V}_0(Q) &=&m_1^2|H_1|^2+m_2^2|H_2|^2+m_3^2
(\epsilon_{ij}{H_1}^i{H_2}^j + h.c.) \nonumber \\
&&+\frac{1}{8}(g^2+g^{\prime 2})\left (|H_1|^2-|H_2|^2\right )^2
+\frac{1}{2}g^2|H_1^{i*}H_2^i|^2\;, \label{eq:3}
\end{eqnarray}
where  $H_1=(H_1^0,H_1^-)$ and
$H_2=(H_2^+,H_2^0)$ are the standard Higgs superfields
and $\epsilon _{ij}$
is the antisymmetric tensor in two dimensions. We have also introduced
the two Higgs mass parameters
\begin{eqnarray}
m_1^2&=&m_{H_1}^2+\mu ^2\label{eq:4}, \\
m_2^2&=&m_{H_2}^2+\mu ^2\label{eq:5}.
\end{eqnarray}
Finally $m_{H_{1,2}}$ and $m_3$ are the soft SUSY breaking mass terms and
$\mu$ is the Higgs mixing mass parameter.

The above tree-level potential ${\cal V}_0(Q)$ depends strongly on the energy
scale $Q$. It has been shown
$^{\cite{GRZ}}$
however, that a correct minimization procedure can be achieved (making
the potential relatively stable), if one includes the one-loop
corrections
$\Delta{\cal V}_1(Q)$
\begin{eqnarray}
\Delta {\cal V}_1(Q)&=&\frac{1}{{64\pi ^2}}{\rm Str}\left [{\cal M}^4
\left (\ln \frac{{{\cal M}^2}}{{Q^2}}-\frac{3}{2}\right )\right ],
 \label{eq:6}
\end{eqnarray}
where ${\cal M}^2$ is the field dependent tree level mass matrix
squared.
Thus finally
\begin{eqnarray}
{\cal V}_H(Q)={\cal V}_0(Q)+\Delta {\cal V}_1(Q)
\end{eqnarray}
The symbol $ Str$  stands for the supertrace which is defined as follows
\begin{equation}
{\rm Str} f({\cal M}^2)=\sum _iq_i(-1)^{2s_i}(2s_i+1)f(m_i^2) \label{eq:7}
\end{equation}
with $q_i$ being the color degrees of freedom while $m_i$ and $s_i$ are the
mass and the spin of the corresponding particle.

Now, electroweak symmetry breaking occurs if the following two conditions are
met:
\begin{itemize}
\item
The supersymmetric Higgs potential should develop an asymmetric
minimum below some scale $Q\le Q_0$. This requirement is expressed
by the condition
\begin{equation}
m_{1}^2(Q)m_{2}^2(Q)-m_3^4(Q) \le 0. \label{eq:9}
\end{equation}
\item
The Higgs potential should be bounded from below. This requirement sets the
second condition, which reads
\begin{equation}
m_{1}^2(Q)+m_{2}^2(Q)\ge 2 |m_3(Q)|^2 \label{eq:10}
\end{equation}
\end{itemize}
The minimization conditions
$\frac{{\partial {\cal V}_H}}{{\partial v_i }}=0$, where
$ v_i \equiv <H_i>$, result the well known equations
\begin{eqnarray}
\frac{1}{2}M_Z^2&=&\frac{{\mu_1^2-\mu_2^2\tan ^2\beta }
}{{\tan ^2\beta -1}}  \label{eq:11} \\
\frac{1}{2}\sin 2\beta&=&-\frac{m_3^2}{\mu_1^2+\mu_2^2}
\label{eq:12}
\end{eqnarray}
where we have introduced the new mass parameters $\mu_i^2=m_{H_i}^2+\mu^2
+\sigma_i^2$, which take into account the corrections to the Higgs
potential from the one-loop contributions $\sigma_i^2$
\begin{eqnarray}
\sigma_i^2\equiv \frac{{\partial \Delta{\cal V}_1}}{{\partial v_i }}
\end{eqnarray}
{}From the above equations one can conclude that the one-loop corrections to
the Higgs potential appear in the minimization conditions through shifts
of the Higgs mass parameters $m_i^2 \rightarrow m_i^2+\sigma_i^2$.
It has been shown
$^{\cite{AN92}}$
, that although $31$ particles contribute to $\sigma_i^2$ corrections,
there are finally large cancellations which reduce significantly their
effect to the electroweak symmetry breaking.
Moreover,
the one-loop contribution of the t-squark-quark sector to the masses
of the
neutralinos, Higgsinos and gauginos seems to be well below the $10\%$
$^{\cite{ppltt}}$
(except in the unfavorable case of a very light tree-level mass)

The most important
contributions arize from the squarks of the third generation and the
top quark mass. Therefore, it is obvious that the Higgs mass
parameters $m_{H_i}$ and the t-squarks play an important role in the
minimization of the Higgs potential.

The scale dependence of $m_{H_i}$ and t-squarks is given
by the renormalization group equations which can be integrated to give
the following results.
The $m_ {H_1}$ Higgs mass parameter is given by
\begin{eqnarray}
m_ {H_1}^2= m_0^2+C_{H_1}(t)m_{1/2}^2
\end{eqnarray}
where $t=\ln Q$, $m_0$ and $m_{1/2}$ are the universal scalar and gaugino mass
parameters at $E_G$, and $C_{H_1}\sim 0.57$ for $t\sim \ln M_Z$. For
the rest of the scalar masses, denoting for convenience  $m_{\tilde
t_L}\equiv \tilde m_1$, $m_{\tilde t_R}\equiv \tilde m_2$ and  $m_
{H_2}\equiv \tilde m_3$, we can write the general analytic
form
$^{\cite{GKL}}$
\begin{eqnarray}
\tilde m^2_n = \alpha_n m^2_0 + C_n(t) m^2_{1/2} - n \delta^2_m(t)
- n \delta^2_A(t)
\label{eq:squarks} \end{eqnarray}
where $\alpha_n$ depends on the K\"{a}ller manifold and hereafter we
assume that $\alpha_n$=1.
The quantities $ \delta^2_{m,A}(t)$ are given by
\begin{eqnarray}
\delta_m^2(t)=\left(\frac{m_t(t)}{2\pi
 v\gamma_Q(t) \sin\beta}\right)^2
\times (3m_0^2I(t)+m_{1/2}^2J(t))\label{eq:dm1}
\end{eqnarray}
and,
\begin{eqnarray}
\delta_A^2(t)&=&\Delta_A^2(t)
-\frac{3}{2}\left(\frac{m_t(t)}{2\pi
 v\gamma_Q(t) \sin\beta}\right)^2E_A^2(t)
\end{eqnarray}
where $v=246$GeV and $I$, $J$, $\Delta^2_A$ and $E_A^2$, are
integrals containing functions of gauge couplings, i.e.
\begin{eqnarray}
I&=&\int_{t}^{t_0}\,dt^{\prime}  \gamma_Q^2(t^{\prime})\\
J&=&\int_{t}^{t_0}\,dt^{\prime}  \gamma_Q^2(t^{\prime})C(t^{\prime})\\
\Delta_A^2&=&\int_{t}^{t_0}
    \frac{h_t^2(t^{\prime})}{8\pi^2}
       A^2(t^{\prime})\,dt^{\prime} \label{eq: Delta}\\
E_A^2&=&\int_{t}^{t_0}\,dt^{\prime}  \gamma_Q^2(t^{\prime})
       \Delta_A^2(t^{\prime})
\label{eq: IJE}
\end{eqnarray}
with $t_0=\ln E_G$, $C(t)\equiv \sum_{n=1}^3C_n(t)$,
 while  $\gamma_Q(t)=
\prod_{j=1}^3 \left(\alpha_{j,0}/\alpha_{j}\right)^{c_Q^j/2b_j}$.
Clearly, a large top mass implies also a large value of the top-Yukawa
coupling, and therefore the negative contributions $\delta^2$
will become also
significant. It is possible then to have
$\tilde m_3^2\equiv m_{H_2}^2$  negative, and the radiative symmetry
breaking scenario will take place.

A very interesting possibily arizes in the case where the top mass is
close to its infrared quasi-fixed  point.
The evolution of the top quark coupling, assuming all the way down
 supersymmetry, is given by
\begin{eqnarray}
h_t &=&\frac{h_t(t_0)\gamma_Q(t)}
{(1 + \frac{3}{4\pi^2} h^2_t(t_0) I(t))^{1/2}}
\label{eq:topeq}
\end{eqnarray}
In this case, i.e.
for $\frac{h^2_t(t_o)}{4\pi} \sim  1$, since $I(t\sim lnm_t)\gg 1$,
we can approximate the above
\begin{eqnarray}
h_t(fixed) \approx \frac
{2\pi}{\sqrt{3I(t)}} \gamma_Q(t)
\label{eq:topfix} \end{eqnarray}
i.e. independent of the initial value $h_t(t_0)$. Thus $m_t(fixed)
=m_t^0\sin\beta \sim (190-200) sin \beta$ GeV, dependening on the
precise values of $\alpha_3$, $E_G$ etc.

The scalar masses of $m_{\tilde{t}_L}$, $m_{\tilde{t}_R}$ and the
Higgs which couples to the up-quarks, take a very simple
$m_t$-independent form in this case. For $h_t=h_t(fixed)$
Eq.(\ref{eq:squarks}) simplifies to
\begin{equation}
\tilde {m}^2_n = (1- \frac {n}{2})
 m^2_0 + \Big[C_n(t) - \frac {n}{6} \frac {J}{I}\Big] m^2_{1/2}
\label{eq:dm1p}
\end{equation}
As far as one  assumes $A(E_G)\le 3|m_0|$,
corrections due to $A$-contributions to the above formula
have been found to be very small and thus they have been totally ignored.
Calculation of the various t-dependent quantities at $t\simeq \ln m_t$ gives
$^{\cite{EKN,GKL}}$
\begin{eqnarray}
C_1(t) \simeq 5.30,\quad
C_2(t)  \simeq 4.90,\quad
C_3 \simeq .57,\quad
I \simeq 113,\quad  J\simeq 590
\label{eq:dm1pp}
\end{eqnarray}

There are some worth-noting properties of the above mass formulae.
Indeed, first note that $ m_{\tilde {t}_R}$ depends only on $ m^2_{1/2}
$ up to $A$ corrections which have been found nigligible.
A second property is that the dependence of the sum
$m^2_{\tilde {t}_L} + m^2_{H_3} $ on $ m^2_0 $, vanishes at the limit
$m_t \rightarrow m_t (fixed)$.
The above properties have also been noticed in
Ref\cite{carena}.
It is interesting to see the implications of the above simplified
formulae in the case of the minimization conditions of the Higgs
potential. We start first with the Higgs mixing parameter $\mu$
involved in the minimization conditions Eqs.(\ref{eq:11}, \ref{eq:12}).
Ignoring one-loop effects for simplicity, the $\mu$ parameter can be
given in terms of the known parameters $I,J$, the unkown Higgs-vev ratio and
the initial values ($m_0,m_{1/2}$), by the following equation
\begin{eqnarray}
|\mu | = \frac{1}{\sqrt{2}} \Big \{\frac{k^2+2}{k^2-1} m_0^2
+ \Big ( \frac{k^2}{k^2-1} \frac{J}{I} - 1\Big ) m^2_{1/2}
 - M^2_Z\Big \}^{1/2}
\label{eq:mu}
\end{eqnarray}
with $k = \tan\beta$.
In Fig.(1),we plot the $|\mu |$-values in the parameter
space ($m_0,m_{1/2}$), for $\tan\beta =1.1$ and $\tan\beta=5$.
For the most of the parameter
space, $|\mu|\le 1.5$TeV. Of course, as $\tan\beta\rightarrow 1$, $\mu$ grows
larger, and a fine tuning problem may arize in Eq.(\ref{eq:11}), in order
to obtain the correct experimetal value of $M_Z$. Thus, to avoid fine
tuning, we may put the condition on $\tan\beta \ge 1.1$, which finally
translates to the bound $M_t\equiv m_t(pole)\ge (150-155)$GeV.

The $\mu$-parameter plays also important role in the squark mass matrices.
In particular, the t-squark mass matrix is
\begin{eqnarray}
M_Q^2&=&\left(\begin{array}{cc}
M_{LL}^2 &M_{LR}^2 \\M_{LR}^2 & M_{RR}^2\end{array}
 \right)
\end{eqnarray}
with eigenvalues given by
\begin{equation}
m^2_{\tilde t_{1,2}}=
\frac{M_{LL}^2+M_{RR}^2\pm
   \sqrt{\left(M_{LL}^2-M_{RR}^2\right)^2+4M_{LR}^4}}{2}
\label{eq:eig}
\end{equation}
where
\begin{eqnarray*}
M_{LL}^2+M_{RR}^2&=&\frac{1}{2}m_0^2
                    +(C_1+C_2-\frac{J}{2I})m_{1/2}^2+2m_t^2+
                      \frac{1}{2}m^2_Zcos 2\beta\\
M_{LL}^2-M_{RR}^2&=&\frac{1}{2}m_0^2
                     +(C_1-C_2+\frac{J}{6I})m_{1/2}^2+
                       (\frac{4}{3}M^2_W-\frac{5}{6}M^2_Z)cos 2\beta\\
M_{LR}^2        &=& m_t^0\left(A\sin\beta+\mu \cos\beta\right)
\end{eqnarray*}
In Fig.(2)  (assuming $\mu > 0$),
we plot contours of the above eigenmasses in the parameter space
($m_0,m_{1/2}$) for the choice $A(E_G)=-\sqrt{3}m_0$, and two representative
values of $\tan\beta$ in the low range $(1.1-10)$, namely
$\tan\beta=1.1$ and $\tan\beta=5$.
In most of the parameter space the light eigenstate preserves the independence
of $m_0$ mass parameter. For reasonable initial values of the parameters
$m_0$ and $m_{1/2}$, the squark masses are well bellow the 1TeV, and therefore
accessible to future experimets.

Notice finally that the one-loop contributions to the effective
potential will also result to a shift in the $|\mu|$ parameter. Making
use of the fact that in the limit $m_{1/2}\gg m_0$
 we can approximate
\[
\ln \frac{m^2_{\tilde t_1}}{M^2_Z}
 \sim \ln \frac{m^2_{\tilde t_2}}{M^2_Z} \sim
\ln \frac{<m^2_{\tilde t}>}{M^2_Z}
\]
we may obtain an analytic form for the one-loop corrected $|\mu|$
parameter, when $m_t$ and $m_{\tilde t_{1,2}}$ corrections are taken
into account
$^{\cite{AN92}}$
\begin{equation}
|\mu|=\sqrt{(\mu _0^2+\eta ^2)/(1-\Omega ^2)}
\label{eq:newmu}
\end{equation}
where
\begin{eqnarray}
\eta ^2&=&\frac{\alpha_2}{8\pi\cos^2\theta_W}
\left\{\left[\left(M^2_{LL}+M^2_{RR}\right) \right.\right.
         \left(\frac{1}{4}-\rho^2\right) \nonumber\\
&+& \left.
\left(M^2_{LL}-M^2_{RR}\right)
          \left(\frac{1}{4}-\frac{2}{3}\sin^2\theta_W\right)
-\rho^2A^2\right]\left(\ln\tilde \rho^2-1\right)\nonumber\\
&-&\left.
2m^2_t\left(\ln\rho^2-1\right)\frac{\rho^2}{k^2}\right\}
\frac{k^2+1}{k^2-1}\\
\Omega^2&=&\frac{\alpha_2}{8\pi\cos^2\theta_W}
\left\{\frac{\rho^2(k^2+1)}{k^2(k^2-1)}\right\}
\left(\ln\tilde \rho^2-1\right)
\label{eq:eta}
\end{eqnarray}
with $\rho=m_t/M_Z$, $\tilde \rho=<m_{\tilde t}>/M_Z$ and $\mu_0$ the
tree level parameter defind in (26). For moderate values of $m_{1/2}$
however, these corrections are not going to alter substantially our
previous results.


\noindent
{\bf Bottom--Tau Yukawa Unification and the IR Fixed Point }

One of the great successes of the most popular GUTs is the equality
of the bottom and tau Yukawa couplings at the GUT scale which lead
to the correct prediction of the experimentally determined relation
$m_{b}\approx 3\,m_{\tau}$ at low energies.
Several groups
$^{\cite{btUN}}$
have examined the effects of $h_b,h_{\tau}$ relations implied by
various unified theories, assuming minimal supersymmetry with grand
unification at an energy scale close to $10^{16}$GeV. It has been
claimed that the GUT relation $h_b=h_{\tau}$  implies a heavy top quark
with a value of the Yukawa coupling close to its infrared fixed point.
In this section, we wish to present a detailed numerical analysis in
the context of the GUT constraints mentioned above. We will mainely
discuss the constraints on the parameter space $(\tan\beta,\alpha_3)$
when bottom-tau Yukawa unification is assumed and examine the
connection of this constraint in relation with the top-mass. We will
further examine the case where the three Yukawa couplings reach the
perturbative limit just after the unification scale. Our analysis will
be done at the two-loop level, taking into account the contribution of
the Yukawa couplings, and in particular that of the $h_t$ into the
running of the gauge coupling constants. Our results largely agree
with previous analyses, however the allowed region in the parameter
space $(\tan\beta,\alpha_3)$ is more constrained. In particular, we
find that  $h_b=h_{\tau}$ can be satisfied only in a small region of
$1\le \tan\beta \le 2$ and  $.111\le \alpha_3 \le .118$.
The case where all three Yukawa couplings are equal at $E_G$ occurs
theoretically in (minimal) $SO(10)$
$^{\cite{ardhs}}$
and in $SU(4)\times SU(2)_L\times SU(2)_R$
$^{\cite{altk}}$
. The allowed region of $\tan\beta$ shortens around the value $63$ for
that case while $\alpha_3(M_Z)$ stays on the lower edge of the
experimentally allowed region ($\sim .11$).

We shall present now a detailed description of the
procedure we are following. We adopt the so called bottom-up approach
starting from $M_Z$. Which are the inputs at this energy level?
\begin{itemize}
\item
the experimentally known values of $\alpha$, $sin^2\theta_W$ and
$\alpha_3$, or equivalently of the three gauge couplings $\alpha_i$,
$i=1,2,3$. The relatively small experimental errors on $\alpha$ and
$sin^2\theta_W$ permit us to talk about the ``bands'' of $\alpha_1$
and $\alpha_2$ in the running of those couplings while we treat
$\alpha_3$ as a ``free'' parameter, inside its experimental limits of
course.
\item
the value of $\tan\beta$, starting its r\^ole
when we reach the energy $E_S$ where SUSY is valid.
\item
the value of the $h_t$ Yukawa coupling of the top quark. Essentially
it is a
free parameter as long as it gives the mass of the top quark in the
allowed experimental region (110--190)GeV. We use the 1 loop QCD
corrections to define the pole mass $M_t$ of the top quark
\[ M_t=\frac{h_t(M_t)v/\sqrt{2}}{1+\frac{4}{3\pi}\alpha_3(M_t)}   \]
\item
the values of $h_b$ and $h_\tau$, taken from the relations
\[ m_b(m_b)=\frac{h_b(M_Z)v/\sqrt{2}}{\eta_b},   \quad\quad
   m_{\tau}=h_{\tau}(M_Z)v/\sqrt{2}\]
We take the mass of the bottom quark $m_b(m_b)=(4.15-4.35)$GeV while
that of the $\tau$ lepton $m_{\tau}(m_{\tau})=1.7841$GeV.
The factor $\eta_b$, appearing in the mass of the bottom quark, includes
the 1 loop QCD corrections from $m_b$ to $M_Z$.
\end{itemize}

Between $M_Z$ and $E_S$, which we take to be 1TeV, we run the
couplings with the $\beta$-functions of the S.M. At $E_S$ we apply the
following boundary conditions for the Yukawa couplings
\[h_t^S=\frac{h_t}{\sin\beta},\quad
  h_b^S=\frac{h_b}{\cos\beta}\quad\mbox{and}\quad
  h_{\tau}^S=\frac{h_{\tau}}{\cos\beta}  \]
Then onwards we run the couplings using the MSSM $\beta$-functions. At
an energy $E_G$, around $10^{16}$GeV, the ``bands'' of  the couplings
$\alpha_1$ and $\alpha_2$ meet and determine what we call ``the
unification band''. The strong coupling $\alpha_3(M_Z)$  should be
chosen so that it passes through this unification band in order to
achieve gauge coupling unification. A short comment is in order at
this point. Since we are using 2 loop $\beta$-functions the
differential equations for all the couplings are coupled (this fact
shows its presence even harder when we demand one or all the Yukawa
couplings to grow large at $E_G$). Therefore the unification band is
not uniquely determined but depends, though not strongly, on the
particular choice for $\alpha_3$ as well as on the Yukawa couplings at
$M_Z$. We try to find the values of $\tan\beta$ that permit the growth
of $h_t$ to the perturbative limit ($h_t\sim 3.5$) at the energy scale
$E_G$ or later, checking always that $\alpha_3$ passes through the
unification band. At the same time we try to unify, again at $E_G$, the
other two Yukawa couplings: $h_b(E_G)=h_{\tau}(E_G)$. This could be
achieved by varying the mass of the bottom quark inside its
experimentally allowed region. Finally we try to arrange the
possibility that all thre Yukawa couplings grow to the perturbative
limit at $E_G$. This last step could be achieved by using large values
of $\tan\beta$.

We approach, step by step, the above three points, constraining in
each step the allowed region of the parameter space of our inputs. In
Fig.3 we plot, for several values of the mass of the top quark $M_t$,
$\tan\beta$ versus $\alpha_3(M_Z)$ demanding gauge coupling unification
and $h_t(E_G)\stackrel{<}{\sim}3.5$. Let us explain the features of
the graph. The lower limit on $\alpha_3(M_Z)$ appears because the
lower the gauge couplings the larger the slope $dh_t/dt$ (recall that
gauge and Yukawa couplings have opposite contributions to the
$\beta$-functions). This fact permits $h_t$ to grow very fast and
reach the perturbative limit before gauge coupling unification is
achieved. The same line of thought explains the slope of the ``lines''
in Fig.3.
Choosing a higher $\alpha_3(M_Z)$ we need a higher initial
point $h_t^S(E_S)$ to reach the perturbative limit, therefore we need a
smaller $\tan\beta$. The turning edges of each line is more intriguing.
At the right end the value of $\alpha_3(M_Z)$ is so high that,
although $h_t$ permits gauge coupling unification, $\alpha_3(E_G)$
passes above the unification band of $(\alpha_1,\alpha_2)$. Choosing a
higher value of $\tan\beta$ (therefore smaller $h_t$) $d\alpha_i/dt$
grows to larger values. The coupling $\alpha_2$ receives the biggest
contribution, the unification band shifts to higher values and
allows $\alpha_3$ to pass through it. Of course, in that case
$h_t(E_G)<3.5$. For each line in Fig.3, the region where
$h_t(E_G)<3.5$ (dashed lines) grows bigger as $m_t$ grows. Similar
arguments explain the left end of the lines. Now $\alpha_3(M_Z)$ is so
small that very easily drops below the unification band. Choosing a
somewhat higher $\alpha_3(M_Z)$ permits a higher $\tan\beta$. Again, in
this case, $h_t(E_G)<3.5$. Therefore the allowed region for each $M_t$
is inside the envelope--like shape.

Our next step is to demand $h_b(E_G)=h_{\tau}(E_G)$. For each $M_t$,
we plot in
Fig.3
a band (shaded region) corresponding to $m_b(m_b)=4.15$GeV (lower line
of the band) and to $m_b(m_b)=4.35$GeV (upper line of the band). We
notice that $b$-$\tau$ unification requires $M_t$ to be near its fixed
point, being closer for $M_t\sim (150-160)$GeV.

The last step is to require all three Yukawa couplings to reach the
perturbative limit near $E_G$. To achieve that point we need a large
value for $\tan\beta$ (for $h_b$ and $h_{\tau}$) and a large $M_t$ (for
$h_t$) The situation is very strict. For example, using as inputs
\begin{eqnarray*}
\tan\beta=63.4,\quad
\alpha_3(M_Z)=.112,\quad
M_t=190{\mbox GeV},\quad{\mbox and} \quad
m_b(m_b)=4.2{\mbox GeV}
\end{eqnarray*}
we get gauge coupling unification at $E_G=(10^{16.0}-10^{16.1})$GeV,
while the three Yukawa couplings at the scale $10^{16.2}$GeV reach
values in the range $3.1-3.5$. Trying to achieve those large values of
$h_t$, $h_b$ and $h_{\tau}$ with a lower value of $M_t$, one needs to
choose either a lower value of $\alpha_3(M_Z)$ or a smaller
$\tan\beta$. The latter does not help since, at such large values,
$\sin\beta$ does not change much while $\cos\beta$ does, preventing
$h_b$ and $h_{\tau}$ to reach the perturbative limit. On the other
hand, the change of $\alpha_3(M_Z)$ does not affect $h_{\tau}$ in
contrast with $h_t$ and $h_b$. The situation is greatly complicated
since all Yukawa couplings are large.

In conclusion, in this paper we examined the implications of a heavy
top quark, and bottom tau unification at the GUT scale, implied by
popular unified models, in the minimal supersymmetric standard model.
We have assumed a top Yukawa coupling close to its infrared fixed
point and we have given analytic forms of the t-squark masses and the
Higgs mass parameter responsible for the radiative electroweak
symmetry breaking scenario. We have found that $m_{\tilde t_L}$ does
not depend on the $m_0$ mass parameter, while all masses under
consideration are very weakly dependent on the trilinear scalar
parameter $A$.
The bottom-tau unification turned out to be very restrictive. We have
found only small ranges in  the $(\tan\beta,\alpha_3$)-plane where
this condition can be satisfied. Moreover, this condition demands a
heavy top with a mass close to its infrared fixed point, as was
previously assumed. In the case of the large $\tan\beta$ scenario, the
above requirements can be satisfied only in a tiny region with
$\tan\beta \approx 63.$

One of us (NDT) would like to thank G. Bathas, K. Farakos, G.
Koutsoumbas and S.D.P. Vlassopulos for useful discussions. The work of
G.K.L. is partially supported by a C.E.C. Science Program
SCI-0221-C(TT), while of N.D.T. by C.E.C. Science Program
SCI-CT91-0729.
\newpage

\newpage

\noindent
{\bf Figure Captions}

\vspace{1cm}
{\bf Fig.1}. Surfaces of constant $|\mu |$ in the parameter space
$(m_0,m_{1/2})$. The upper surface corresponds to $\tan\beta=1.1$
while the lower one corresponds to $\tan\beta=5$.

{\bf Fig.2}. Contours of constant $m_{\tilde t_{1,2}}$ in the
parameter space $(m_0,m_{1/2})$, for two values of $\tan\beta=1.1$ and
$5$. (a) and (c) corresponds to the lighter eigenstate while (b) and (d)
to the heavier one.

{\bf Fig.3}. Allowed regions in the space of ($\tan\beta,
\alpha_3(M_Z)$), in order to achieve gauge coupling unification, for
several values of the top mass $M_t$. Below the solid part of the
contours, $h_t$ reaches the perturbative limit before gauge coupling
unification, while below the dashed part $\alpha_3$ passes above the
unification band of $\alpha_1$ and $\alpha_2$. Demanding
$h_b=h_{\tau}$ at $E_G$, the allowed regions, for each $M_t$ value,
shrink to the corresponding shaded bands.


\begin{thebibliography}{99}
\bibitem{IR}L. Ib\'a\~nez and G.G. Ross, Phys. Lett. 110B(1982)215.
\bibitem{RAD}
K. Inoue et al, Prog. Theor. Phys. 68(1982)927;\\
L. Alvarez-Gaum\'e, M. Claudson and M.B. Wise, Nucl. Phys.
B221(1983)495;\\
J. Ellis, J. S. Hagelin and K. Tamvakis, Phys. Lett. B125(1983)275;\\
L. Ib\'a\~nez and C. Lopez, Phys. Lett. B126(1983)54;\\
C. Kounnas, A. B. Lahanas, D.V. Nanopoulos, and M. Quir\'os,
Phys. Lett. B132(1983)95; Nucl. Phys. B236(1984)438.
\bibitem{PR}B. Pendleton and G.G. Ross, Phys. Lett. 98B(1981)291.
\bibitem{GRZ}G. Gamberini, G. Ridolfi and F. Zwirner, Nucl. Phys.
B331(1990)331.
\bibitem{AN92}R. Arnowitt and P. Nath, Phys. Rev. D46(1992)3981.
\bibitem{ppltt}D. Pierce and A. Papadopoulos, John Hopkins University
preprints, JHU-TIPAC-930030;JHU-TIPAC-940001;\\
A.B. Lahanas, K. Tamvakis and N.D. Tracas, CERN preprint, CERN-TH
7089/93 (Phys. Lett. B, to appear).
\bibitem{GKL}G.K. Leontaris, Phys.Lett.B317(1993)569.
\bibitem{EKN}
C. Kounnas, A. B. Lahanas, D.V. Nanopoulos, and M. Quir\'os,
Phys. Lett. B132(1983)95; Nucl. Phys. B236(1984)438;
J. Ellis, C. Kounnas and D. V. Nanopoulos, Nucl. Phys.
B241(1984)406.
\bibitem{carena}
W.A. Bardeen, M. Carena, S. Pokorski and C.E.M. Wagner, MPI preprint
MPI-Ph/93-58;\\
M. Carena, M. Olechowski, S. Pokorski and C.E.M. Wagner, MPI
preprints, MPI-Ph/93-66 and MPI-Ph/93-94.
\bibitem{btUN}G.K. Leontaris, J. Rizos and  K. Tamvakis, Phys. Lett. B
243(1990)220;\\
B. Ananthanarayan, G. Lazarides and Q. Shafi, Phys. Rev.
D44(1991)1613;\\
H. Arason et al, Phys. Rev. Lett. 67(1991)2933;\\
V. Barger, M.S. Berger and P. Ohmann, Phys. Rev. D47(1993)1093;\\
G.L. Kane et al, UM-TH-93-24, October 1993;\\
G.K. Leontaris nad N.D. Tracas, Phys. Lett. 303B(1993)50;\\
J. Lopez, D.V. Nanopoulos and A. Zichichi, CERN preprint CERN-TH.
7138/94;\\
B Ananthanarayan, K.S. Babu and Q. Shafi, Bartol Research Institute
preprint, BA-94-02;\\
P. Langacker and N. Polonsky, February 1994,UPR-0594T  preprint;\\
B.C. Allanach and S. King, preprint SHEP-93/94-15.
\bibitem{ardhs}G. Andersom, S. Raby, S. Dimopoulos, L. Hall and G.
Starkman, LBL preprint 33531.
\bibitem{altk}I. Antoniadis et al, Phys. Lett.
216B(1989)333;245(1990)161;\\
G.K. Leontaris and N.D. Tracas, Phys. Lett. 260B(1991)339;\\
A Murayama and A. Toon, Phys. Lett. B318(1993)298;\\
O. Korakianitis and N.D. Tracas, Phys. Lett. 319B(1993)145;\\
S. King, Southampton Preprint, SHEP 93/94-08 (Phys. Lett. B, to
appear),\\
G. Lazarides and C. Panagiotakopoulos, Universiy of Thessaloniki
preprint, 3/1994.
\end{thebibliography}
\end{document}